\documentclass[aps,twocolumn,superscriptaddress,notitlepage,longbibliography]{revtex4-1}%
\usepackage{float}
\usepackage{graphicx}
\usepackage{amsmath,amssymb,bbm,mathrsfs,bm,braket,color,graphicx,comment,amsfonts,dsfont}
\usepackage[colorlinks,citecolor=blue,urlcolor=blue]{hyperref}
\usepackage[mathscr]{euscript}
\usepackage[normalem]{ulem}
\usepackage{comment}
\DeclareMathAlphabet\mathbfcal{OMS}{cmsy}{b}{n}

\usepackage{extarrows} 
\usepackage{mathrsfs}

\definecolor{purple}{rgb}{0., 0.0, 1.}
\definecolor{grun}{rgb}{0.2, 0.6, 0.5}

\usepackage{tikz,lipsum,lmodern}
\usepackage[most]{tcolorbox}

\begin{document}

\title{Noise resilience in adaptive and symmetric monitored quantum circuits}


\author{Moein N. Ivaki}
\email{moein.najafiivaki@aalto.fi}
\address{InstituteQ and QTF Centre of Excellence, Department of Applied Physics,
School of Science, Aalto University, FI-00076 Aalto, Finland}

\author{Teemu Ojanen}
\address{Computational Physics Laboratory, Physics Unit, Faculty of Engineering and Natural Sciences, Tampere University, P.O. Box 692, FI-33014 Tampere, Finland}
\address{Helsinki Institute of Physics P.O. Box 64, FI-00014, Finland}

\author{Ali G. Moghaddam}\email{a.ghorbanzade@gmail.com}
\address{Computational Physics Laboratory, Physics Unit, Faculty of Engineering and Natural Sciences, Tampere University, P.O. Box 692, FI-33014 Tampere, Finland}
\address
{Helsinki Institute of Physics P.O. Box 64, FI-00014, Finland}
\address{Department of Physics, Institute for Advanced Studies in Basic Sciences (IASBS), Zanjan 45137-66731, Iran}

\date{\today}

\begin{abstract}
Monitored quantum circuits offer great perspectives for exploring the interplay of quantum information and complex quantum dynamics. These systems could realize the extensively-studied entanglement and purification phase transitions, as well as a rich variety of symmetry-protected and ordered non-equilibrium phases. The central question regarding such phases is whether they survive in real-world devices exhibiting unavoidable symmetry-breaking noise. We study the fate of the symmetry-protected absorbing state and charge-sharpening transitions in the presence of symmetry-breaking noise, and establish that the net effect of noise results in coherent and incoherent symmetry-breaking effects. The coherent contribution removes a sharp distinction between different phases and renders phase transitions to crossovers. Nevertheless, states far away from the original phase boundaries retain their essential character. In fact, corrective feedback in adaptive quantum circuits and postselected measurements in symmetric charge-conserving quantum circuits can suppress the effects of noise, thereby stabilizing the absorbing and charge-sharp phases, respectively.
Despite the unavoidable noise in current quantum hardwares, our findings offer an optimistic outlook for observing symmetry-protected phases in currently available Noisy Intermediate-Scale Quantum (NISQ) devices. Moreover, our work suggests a symmetry-based benchmarking method as an alternative for characterizing noise and evaluating average local gate fidelity.\end{abstract}

\maketitle

\section{Introduction}
Understanding the relationship between quantum dynamics and information flow in complex far-from-equilibrium many-body systems has become a major theme in contemporary research. In particular, the interplay of unitary evolution, which drives entanglement growth and information scrambling, and \emph{non-unitary} monitoring by local measurements, which suppresses entanglement proliferation, has stimulated a flurry of recent activity~\cite{li_quantum_2018,skinner_measurement-induced_2019,choi_quantum_2020,gullans_dynamical_2020}. This interplay has given rise to an entirely new paradigm, characterized by the fascinating phenomena known as \emph{measurement-induced phase transitions} (MIPTs)~\cite{potter_entanglement_2021,Fisher_random_quantum_circuits,bao_theory_2020,li_measurement-driven_2019,Jian-Ludwig2020_measurement_criticality,ippoliti_entanglement_2021}. A compelling approach to enhance the observability of MIPTs involves guiding the dynamics of a quantum system through the implementation of additional feedback gates, conditioned on measurement outcomes~\cite{li2021robust, PhysRevLett.129.200602, Li_crossentropy, garratt2022measurements, buchhold2022revealing,Iadecola_dynamical,McGinley_feedback_dtc}. A typical example of this approach is the employment of interactive feedbacks based on measurement results, to direct the system towards \emph{absorbing states}~\cite{Ravindranath2023,Khemani2022Absorbing, hauser2023continuous, PhysRevLett.131.060403, PhysRevLett.130.120402}, which typically exhibit symmetry breaking of a global symmetry~\cite{PhysRevLett.116.245701, PhysRevE.94.012138, PhysRevA.96.041602, lesanovsky2019non, PhysRevLett.123.100604, chertkov2022characterizing}. Moreover, a wide variety of predicted phases and phenomena arising from the monitored dynamics, such as the charge-sharpening transition~\cite{Agrawal_2022, Chakraborty2024}, rely on the presence of specific symmetries of the unitaries. However, the impact of symmetry-breaking, which is inevitable in real quantum hardware, remains largely unexplored.

\begin{figure}[!]
\centering
\includegraphics[width=.99\linewidth]{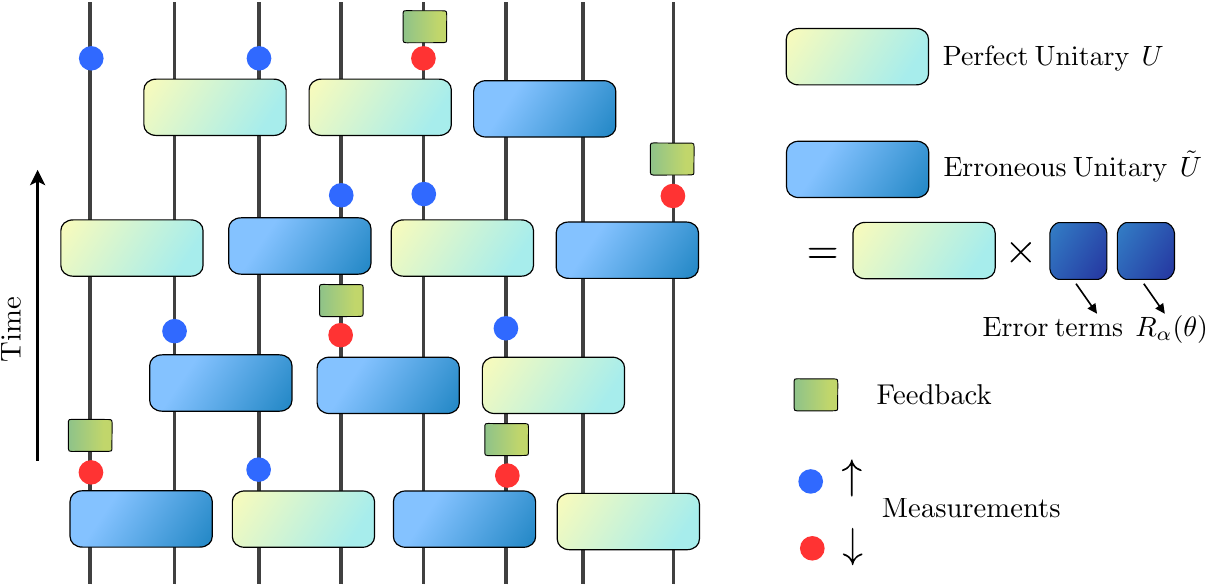}
\caption{\label{fig1} 
\textbf{Monitored adaptive quantum circuits with noise}:
The schematic illustrates a monitored quantum circuit with single-qubit corrective feedback (green squares) that directs the system toward a fully polarized absorbing state in the presence of generic single-qubit errors. The effect of single-qubit noise can be absorbed into the two-qubit unitaries (large rectangles), resulting in gates that are no longer ideal but contain some level of error. While perfect unitaries preserve certain symmetry constraints of an ideal circuit model, the erroneous gates can generally break those symmetries.
}
\end{figure}

In this work, we address the effects of symmetry-breaking single qubit gates noise two types of monitored quantum circuits: (i) adaptive circuits with corrective feedback that steers the system toward an ordered (absorbing) state (Fig.~\ref{fig1}); and (ii) non-adaptive  $U(1)$-symmetric monitored circuits, which lack feedback but utilize $U(1)$ charge conserving unitaries.
Errors in the execution of quantum gates are a major source of noise in current digital quantum simulators and vast efforts have been devoted to quantum error correction and error mitigation techniques
\cite{Emerson2015,Flammia2016PRL,Greenbaum_2018,bravyi2018correcting,Preskill_2020,Flammia_2019,Wallraff_2020}. 
These random errors, which do not respect symmetries of the circuit, pose a particular challenge for symmetry-protected phenomena. Moreover, due to the irreproducible nature of noise across different runs of the same circuit, the ensemble resulting from multiple state preparations no longer represents a pure state but rather a statistical mixture.

Our key finding is that the net effect of symmetry-breaking noise gives rise to coherent and incoherent symmetry-breaking contributions. The coherent contribution can be regarded as an additional symmetry-breaking contribution to the symmetry-preserving unitary gates, thus removing the strict distinction of symmetry-protected phases and transforming the phase boundaries into crossovers. Despite this, states far away from the original phase boundaries retain their essential nature even in the presence of strong symmetry-breaking noise. This enables observation of weakly-modified absorbing states and sharp charge states at higher monitoring rates.  In adaptive circuits, the higher monitoring rate is relatively cost-free as the absorbing state transition does not require postselection~\cite{Khemani2022Absorbing}. However, the higher monitoring rate to observe a charge sharpening in $U(1)$-symmetric quantum circuits comes with a higher postselection overhead.

\section{Results}

\emph{\color{blue} Noise model}.---We start with error-free monitored quantum circuits, where the Haar-random two-qubit unitaries are constrained by specific symmetry operators. These circuits can be adaptive, meaning that measurements are followed by corrective feedback unitaries conditioned on the measurement outcomes~\cite{Ravindranath2023,Khemani2022Absorbing}. We also consider non-adaptive charge-conserving $U(1)$ circuits, which exhibit a charge-sharpening transition~\cite{Agrawal_2022}.

Then, we introduce generic random single-qubit rotation gates of the form $R^{j}_{\alpha}(\theta_j)=e^{i\theta_j \sigma_{\alpha}/2}$, with \(\alpha\in\{x,y,z\}\), \(j\in [1,L]\) and $L$ the number of qubits~\cite{bravyi2018correcting,Emerson2015}. These additional single-qubit gates, which probabilistically act on an \emph{extensive} number of randomly chosen qubits at each time step, introduce symmetry-breaking noise into the circuit. 
The advantage of defining errors as rotation operators is that the symmetry-breaking effect can be tuned continuously by varying the range of rotation angles.
For each qubit, the noise angle and rotation direction (indicated by $\alpha$) are chosen randomly with \(\theta_j\) lying in the interval \([0,\pi\Theta]\), where \(0\leq \Theta \leq 1\) determines the maximal noise amplitude. For the sake of concreteness,
we assume that the noise terms always follow perfect unitaries 
\(U^{j,j+1}\) making them effectively imperfect \(\tilde{U}^{j,j+1}\equiv U^{j,j+1}V^{j,j+1}\), with \(V^{j,j+1} =R_{\alpha}^{j}(\theta_{j})\otimes R_{\alpha'}^{j+1}(\theta_{j+1})\)~\cite{footnote1}. 

The average \textit{error strength} depends both on the average ratio of faulty unitaries at each time step, which can be denoted as noise rate $\gamma$, as well as on the maximum rotation angle $\pi\Theta$. In the context of noisy quantum channels, the strength of noise is commonly described using the average gate fidelity, which is related to the noise amplitude $\Theta$ as detailed in in App.~\ref{app:gate-fidelity}. We show that the average gate fidelity for single-qubit gates is
given by
\begin{align} \label{eq:fiedelity-noise-relation}
        {\mathbb F}_{\rm ave} (\Theta,\gamma) =  1 - \frac{\gamma }{3}\left( 1-\frac{\sin (\pi\Theta)}{\pi \Theta}\right),
\end{align}
assuming a noise rate $\gamma\in[0,1]$ 
and noise amplitude $\Theta \in[0,1]$.
When either of the noise rate or its amplitude increase, the gate fidelity decreases, with the minimum value of $2/3$ corresponding to $\gamma=1$ and $\Theta=1$. In our setup, the noise rate is fixed to $\gamma=1/2$, and we primarily focus on the role of $\Theta$. However, all of our results can equally be interpreted in terms of average gate fidelity, as shown in Eq.~\eqref{eq:fiedelity-noise-relation}.

It should be emphasized that we only consider noise that can be described as a set of independent single-qubit errors occurring during circuit execution. In reality, there are also correlated two-qubit and multi-qubit errors that cannot be decomposed into a direct product of single-qubit errors. However, the probability of such errors affecting multiple qubits simultaneously typically decreases exponentially as $\epsilon^{-m}$, where $m$ is the number of correlated qubits involved. Here, $\epsilon$ represents the single-qubit error rate. This behavior is a key characteristic of noise and errors generated by a sufficiently random process, and it forms the foundation for why, for instance, \emph{quantum error correction} can work in practice.

Another point about the noise model introduced here is that, while single-qubit gate errors do not affect coherence in a single execution of the circuit, they give rise to decoherence when averaged over multiple runs. In multiple executions, these coherent errors result in an incoherent contribution, leading to a noisy quantum channel. As shown in App.~\ref{app-channels}, the quantum channel resulting from averaging over many realizations of single-qubit gate errors can be decomposed into coherent and incoherent components. The coherent contribution takes the form of a unitary transformation
\begin{equation}
  \rho \to \rho' = {\cal E}^{\rm coh}_{\alpha}(\rho) = R_{\alpha}(\varphi) \rho R_{\alpha}^\dagger(\varphi),
\end{equation}
while the incoherent contribution is expressed in terms of generic Pauli depolarization channels
\begin{align}
    \rho \to \rho' = {\cal E}^{\rm incoh}_{\alpha} (\rho) = (1-\eta) \rho + \eta\: \sigma_\alpha \rho \sigma_\alpha.
\end{align}
The parameters determining the coherent and incoherent part of quantum channels are found as $\varphi = (\pi/2)\Theta$ and $\eta = 1/2 - \sin(\pi\Theta/2)/(\pi\Theta)$ as functions of the noise amplitude $\Theta$.

The fully incoherent part of the noise channel, ${\cal E}^{\rm incoh}$, generates a mixed state. In both numerical simulations and real experiments, this occurs by executing the quantum circuit many times, each time with a randomly different set of rotations $\{R_{\alpha}^{j}(\theta_j)\}$. While this incoherent noise readily does not respect the symmetry, the coherent contribution, ${\cal E}^{\rm coh}$, more explicitly breaks the symmetry by introducing the preferential rotations with average angles $\phi = (\pi/2)\Theta$. As we will see, this explicit symmetry breaking plays a crucial role in smoothing out the phase boundaries between fuzzy and ordered states into crossovers.

\emph{\color{blue} Absorbing state transition}.---The absorbing phases are realized by driving the system into a desired ordered state through a combination of certain unitary transformations and local corrective feedback operators. Here, we consider a model where the target state is the trivial fully-polarized state $\psi_{\rm tar} = \ket{111 \cdots 1}$. In this case, the two-qubit unitary operators take the symmetric form
\begin{align}
 U^{j,j+1} =  \begin{pmatrix}
 U_{3 \times 3}  &  \\
 & e^{i\phi}
    \end{pmatrix},
    \label{eq:unitary_adaptive}
\end{align}
in the basis $\{\ket{00}, \ket{01}, \ket{10}, \ket{11}\}$, leaving the $\ket{11}$ subspace intact. Each measurement is then followed by a local feedback, applying a bit-flip operation $R_x(\pi) \equiv \sigma_x$ to correct any undesired measurement outcome $\ket{0}$.

Adaptive circuits typically host three distinct phases (i) non-absorbing volume-law,
(ii) non-absorbing area-law, and
(iii) absorbing area-law
separated by two transition points $p_{\rm c}$ and $p_{\rm abs}$. For high enough measurement rates $p> p_{\rm abs}\gtrsim p_{\rm c}$
the circuit undergoes an absorbing-state transition characterized by an order parameter $\bar{n} = 2 \langle Q\rangle/L -1 $~\cite{footnote2}.
The total charge operator $Q=\sum_j q_j $ in terms of single-qubit charge operators $q_j = (1+Z_j)/2 $ such that $q_j\ket{1}_j=\ket{1}$ and  $q_j\ket{0}_j=0$.
Absorbing and fuzzy state thus correspond to 
$\bar{n}=1$ and $\bar{n}=0$, respectively.
A key characteristic of absorbing states is their exponentially fast convergence over time, with timescales $t_{\rm abs} \lesssim L$, in contrast to the fuzzy regime~\cite{Khemani2022Absorbing}.

An interesting feature of absorbing-state transitions and adaptive circuits is that feedback mechanisms, which steer the system toward a macroscopically ordered state, inherently eliminate the need for postselection~\cite{buchhold2022revealing, Sierant_Turkeshi_2023}. This can be understood by recognizing that the average of postselected results for the order parameter $\bar{n}$ is equivalent to the order parameter associated with the mixed state given by the (non-postselected) average density matrix $\rho$, due to the linearity of order parameter in $\rho$. This property, where the averages over expectation values and density matrices match, naturally persists even in the presence of noise. Therefore, postselection for both measurement outcomes and noise operators becomes unnecessary in adaptive quantum circuits to reveal the absorbing-state transition~\cite{PRXQuantum.4.040309}. However, although the symmetry of applied gates and corrective feedback ensures the transition to the absorbing state in the absence of noise, it is not immediately evident whether the symmetry-breaking effects of noise can be significantly compensated without resorting to postselection. This issue will be discussed next.

\begin{figure}[htp]
\centering
\includegraphics[width=0.9\linewidth]{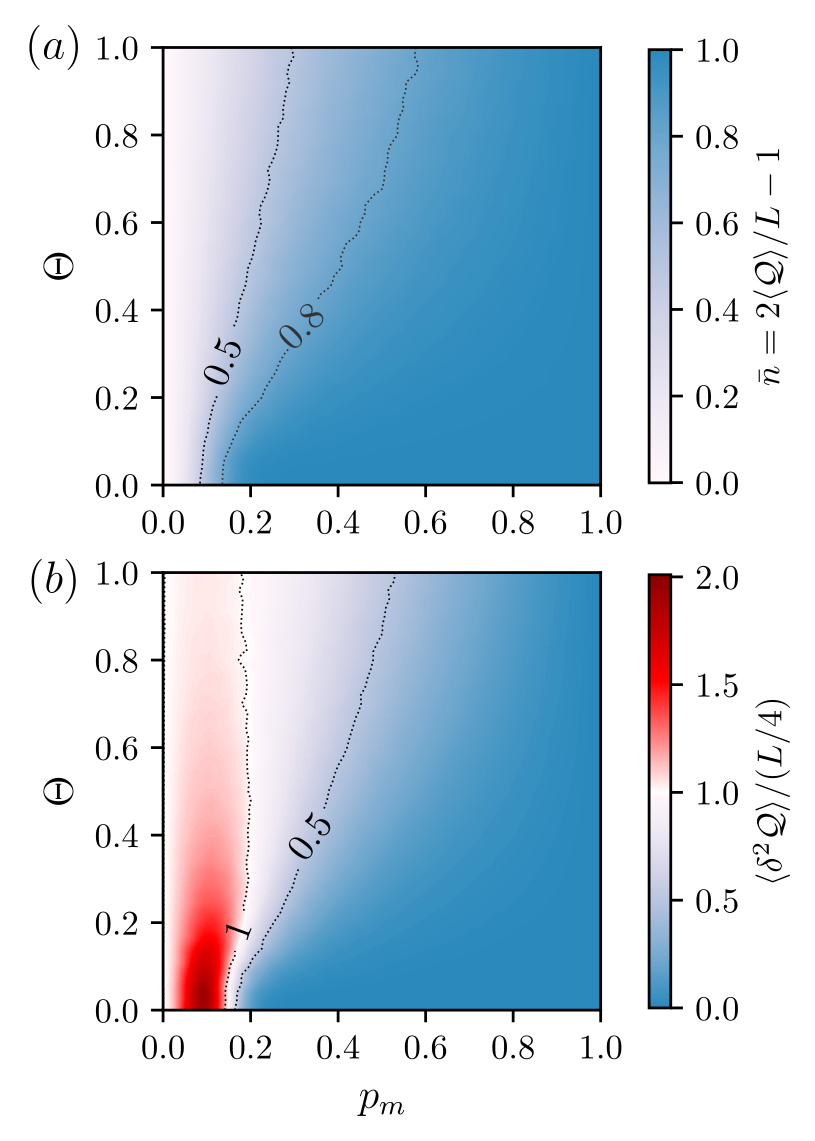}
\caption{\label{fig:heatmaps_adaptive} 
    \textbf{Noise effect on adaptive monitored circuits.} 
    \textbf{(a)} Absorbing-states order parameter $\bar{n}=2\langle {\cal Q} \rangle/L -1$ as a function of measurement rate $p_{\rm m}$ and noise amplitude ${\Theta}$.
    The order paramater varies between zero and unity corresponding to completely fuzzy and fully ordered states which occur for $p_{\rm m}=0,1$, respectively. For any constant noise amplitude, by increasing the measurement rate, the order parameter increases and system reaches its long-time steady state which is a noisy absorbing state with $\bar{n}\lesssim 1$. \textbf{(b)} Scaled charge fluctuations are relatively large and small in fuzzy and ordered regimes, respectively. The transition/crossover between the two regions roughly corresponds to $\langle \delta^2{\cal Q} \rangle/(L/4)\approx 1$. 
    Both quantities are calculated for mixed states with the density matrix averaged over
    1000 realizations of a circuit with $L=12$ and time averaging over $3L<t<4L$. Particularly, we should mention that there is no postselection over measurement outcomes even for fluctuations.
}
\end{figure}

\emph{\color{blue} Noisy adaptive circuits}.---As discussed earlier, we model the noise by random single-qubit rotation gates on top of the symmetric unitaries defined in Eq. \eqref{eq:unitary_adaptive}. These additional operations due to the noise violate the symmetry of perfect unitaries because they can connect the $\ket{11}$ state to other sectors, namely $\ket{10}$ and $\ket{01}$.
The symmetry of perfect unitaries 
can be described by the operator
\begin{align} \label{eq:symm-operator-abs}
{\Pi}_{\rm abs} = {\rm diag}(1,1,1,0),
\end{align}
which implies that $[U^{j,j+1},{\Pi}_{\rm abs}]=0$. 
Considering noisy unitaries, it can be easily verified that the noise component, $V^{j,j+1}$, which consists of two independent single-qubit rotations, generally does not commute with the symmetry operator \eqref{eq:symm-operator-abs}.
In adaptive circuits, it is important to note that feedback corrections can also disrupt the symmetry of unitaries. However, for any pair of qubits that have successfully reached the target state through local operations, they become \emph{frozen}, and the symmetry is preserved for this specific pair due to the conditional nature of the corrective feedback. This is because noiseless two-qubit unitaries and subsequent measurements will not alter the target state, eliminating the need for corrections that could otherwise break the symmetry. Conversely, noisy unitaries always explicitly break the symmetry.

To explore the effects of noise, we begin by analyzing the behavior of the averaged order parameter $\bar{n}$ and the charge fluctuations $\langle\delta^2{\cal Q}\rangle = \langle {\cal Q}^2 \rangle - \langle {\cal Q} \rangle^2$~\cite{PhysRevLett.131.020401} as functions of the measurement rate $p_{\rm m}$ and the average noise amplitude $\Theta$, as shown in Fig.~\ref{fig:heatmaps_adaptive}. The density matrix, over which these average values are defined, is given by 
\begin{align}
\rho(t) &= \mathbbm{E}_{\cal R,M} 
(\ket{\Psi_{\cal R,M}(t)}\bra{\Psi_{\cal R,M}(t)})
\nonumber\\
&\approx \frac{1}{{\cal N}_{\rm runs}} \sum_{i=1}^{{\cal N}_{\rm runs}} \ket{\Psi_i(t)}\bra{\Psi_i(t)}.
\label{eq:rho_adaptive}
\end{align}
Here, each individual run of the circuit involves both random choices of noise terms ${\cal R} = \{R^j_{\alpha}(\theta_j)(t)\}$ and possible measurement outcomes ${\cal M} = \{m_j(t)\}$ throughout the circuit evolution (where $j$ denotes the qubit index and $\alpha$ the rotation axis of the noise term). Essentially, the system evolves under a noisy quantum channel, resulting in the mixed state $\rho$, due to the probabilistic nature of both the noise and the measurement outcomes. The results obtained using the density matrix $\rho$ represent non-postselected, noise-averaged quantities, which are also straightforward to measure in practical experiments. It is important to note that the first line of Eq.~\eqref{eq:rho_adaptive} provides the exact theoretical definition of the density matrix corresponding to the noisy, non-postselected channel, while the second line presents an approximate form based on finite sampling in numerical simulations, and naturally, in real experiments.

Without monitoring ($p_{\rm m} = 0$), the circuit approaches a completely fuzzy state with $\bar{n} = 0$, regardless of the noise level. Conversely, perfect monitoring ($p_{\rm m} = 1$) leads to a fully polarized state with $\bar{n} = 1$. Varying the measurement rate at a constant noise amplitude results in an increase in the order parameter, eventually reaching a noisy absorbing state with $\bar{n} \lesssim 1$. At low measurement rates, a fuzzy phase persists, characterized by $\bar{n} \lesssim {\cal O}(1/L)$, where at most ${\cal{O}}(1)$ qubits are polarized. Hence, in the limit of large circuits, $\bar{n}$ approaches zero as ${\cal O}(1/L)$. The noisy absorbing-state regime, in contrast, corresponds to a region where a finite fraction of all qubits are ordered. As observed in Fig.~\ref{fig:heatmaps_adaptive} (a), the transition from the fuzzy to the absorbing phase occurs around $p_{\rm m} \approx 0.1$ in the absence of noise, consistent with previous studies. Increasing the noise amplitude necessitates a higher measurement rate to achieve this transition, as evidenced by the shape of the constant $\bar{n}$ lines. In essence, the presence of noise requires a higher measurement rate to obtain a specific steady-state value of the order parameter.

The properties of the absorbing and fuzzy regimes in the presence of noise are also reflected in the charge fluctuations, as shown in Fig.~\ref{fig:heatmaps_adaptive} (b). In the absence of noise, the absorbing state exhibits vanishingly small fluctuations, while the fuzzy phase displays finite fluctuations that follow a volume-law scaling. At $p_{\rm m} = 0$, the fluctuations reach $\delta^2{\cal Q} = L/4$, which is the value for a fully mixed state where each qubit has an equal probability of being in $\ket{0}$ or $\ket{1}$. Introducing measurements initially increases the fluctuations, and after reaching a peak around the transition point, they begin to decrease as the absorbing state regime is achieved ($p_{\rm m} > p_{\rm abs}$). This overshooting behavior indicates the system's failure to effectively transition toward the absorbing state at low $p_{\rm m}$. In this regime, the combination of a low measurement rate, feedforward corrections, and random unitary dynamics generates even more fluctuations. Figure~\ref{fig:heatmaps_adaptive} (b) suggests that this characteristic persists in the presence of noise, although it weakens and eventually almost disappears at very high noise amplitudes.

\begin{figure}[!]
\centering
\includegraphics[width=0.9\linewidth]{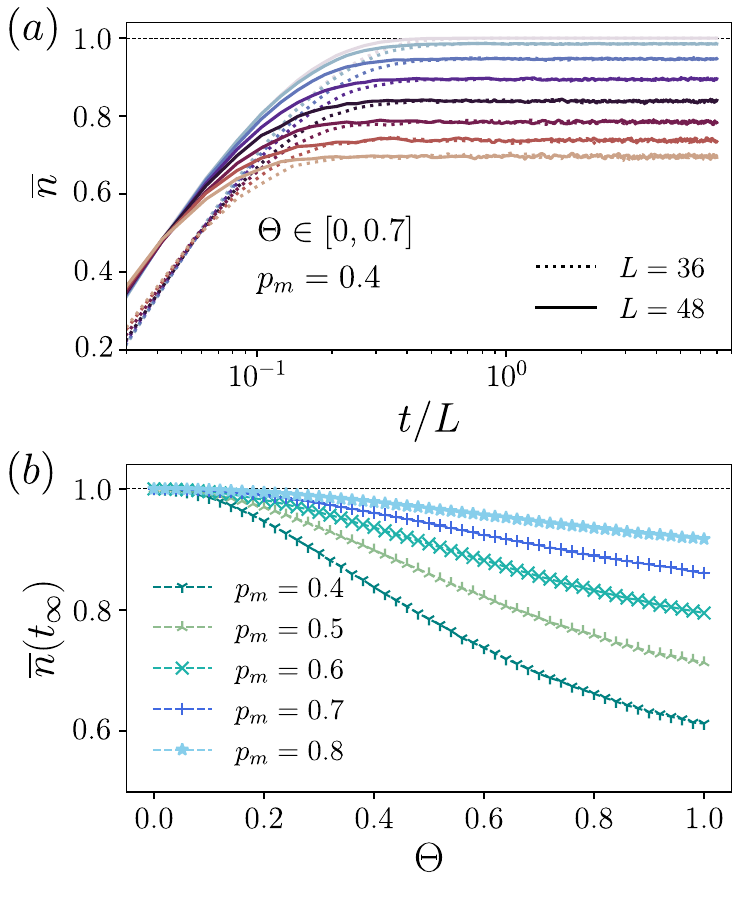}
\caption{\label{fig3} 
{\textbf{Properties of absorbing states in noisy circuits utilizing MPS.} {\textbf{(a)}} Dynamical behavior of the order parameter as a function of $t/L$. {\textbf{(b)}} Long-time behavior of $\bar{n}$ for different measurement rates $p_{\rm m}$, plotted as a function of the maximum noise amplitude $\Theta$}.}
\end{figure}

Fluctuations provide an additional tool for distinguishing between the fuzzy and absorbing states in the presence of noise. In the fuzzy state, fluctuations are equal to or larger than $\delta^2{\cal Q} = L/4$, corresponding to a fully mixed state where each qubit has equal probabilities of being in $\ket{0}$ or $\ket{1}$. Conversely, the absorbing state exhibits small fluctuations, with $\delta^2{\cal Q} \ll L/4$. This suggests that while there is a sharp transition between the two phases, marked by an abrupt change in fluctuations at very low noise levels, finite noise causes this transition to evolve into more of a crossover behavior. As a result, the extent of this crossover region, in terms of measurement rate, grows as the noise amplitude increases. This is illustrated by the two hypothetical threshold values $\delta^2{\cal Q} / (L/4) = 1, 0.5$ in Fig.~\ref{fig:heatmaps_adaptive} (b). From another perspective, this crossover region might be associated with the expansion of a small area-law non-absorbing state, which exists in the absence of noise~\cite{Khemani2022Absorbing}, into the domain of noisy adaptive circuits.

To further illustrate the noise resilience of absorbing states, we investigate the behavior of the order parameter $n_d$ at intermediate and high measurement rates for larger system sizes as shown in Fig.~\ref{fig3}. At such measurement rates the weak entanglement allows us to employ a matrix product state (MPS) approach, enabling us to reach system sizes up to 48 qubits~\cite{orus2014practical, pastaq}. A key characteristic of absorbing states, in the absence of noise, is their exponentially fast convergence over time, with timescales $t_{\rm abs} \lesssim L$~\cite{Khemani2022Absorbing}. Interestingly, our calculations indicate that even in the presence of strong noise, this behavior persists at sufficiently high measurement rates, as illustrated in Fig.~\ref{fig3}(a). Specifically, at a measurement rate of $p_{\rm m} = 0.4$, the overall time evolution of the order parameter remains unchanged, with only the steady-state value being influenced by the noise amplitude, as anticipated. 
Furthermore, we observe that the steady-state value ${\bar n}(t_{\infty})$ is independent of the system length, strongly suggesting that this behavior persists for thermodynamically large systems. Figure~\ref{fig3}(b) demonstrates the dependence of ${\bar n}(t_{\infty})$ on noise amplitude for various measurement rates. Increasing the noise amplitude leads to a noisy absorbing state characterized by smaller long-time values of the order parameter. However, as mentioned earlier, increasing $p_{\rm m}$ can mitigate the effect of the noise.

Our findings are similar in spirit to those reported in Ref.~\cite{PhysRevA.96.041602}, where the transition evolves into a crossover, or in the classical picture, the nature of the transition changes. This behavior is closely tied to the partially coherent nature of the noise discussed earlier, which introduces fully quantum dynamical effects. In fact, without the coherent contribution, the absorbing transition can be effectively described as a classical process~\cite{Khemani2022Absorbing}. However, in the presence of noise, as discussed in App.~\ref{app-measurement-noise-channel}, the coherent part leads to non-classical behavior that can alter the universality class of the transition~\cite{PhysRevB.95.014308, PhysRevLett.116.245701}. A detailed study of the universality classes governing the transition or crossover between the absorbing and non-absorbing phases will be addressed in future work.

\emph{\color{blue} Charge-sharpening transition}.---Monitored random quantum circuits with global symmetries, namely a $U(1)$ charge conservation, exhibit a distinct transition in addition to the entanglement transition, known as the \emph{charge-sharpening} transition. This transition can be best understood as a coarse-grained form of purification transitions~\cite{footnote3}.
The main difference between charge-sharpened states and purified states is that it is sufficient to have a fixed charge for the former, whereas the latter requires purification to a specific many-body state~\cite{ippoliti2023learnability}. 
Since the charge-sharp phase occurs when the system dynamics purifies it to a certain charge sector, it can still be mixed within the charge sector. 
This shows that the charge-sharpening transition occurs slightly earlier than the entanglement transition ($p_{\#}<p_{\rm c}$).

Two-qubit unitaries in these circuits follow the form
\begin{align}
     U^{j,j+1} =  \begin{pmatrix}
     e^{i\phi_{00}} &  & \\
     &U_{2 \times 2}&  \\
     &  & e^{i\phi_{11}}
    \end{pmatrix},
\end{align}
possessing charge conservation described by the symmetry operator
\begin{align}
    \Pi_{ U(1)} = {\rm diag}(0,1,1,2),
\end{align}
such that $[U^{j,j+1},    \Pi_{ U(1)}]=0$
To study the charge-sharpening transition, we feed the circuit with an initial state prepared in an equal superposition of all possible charge sectors $\otimes_{i=1}^{L}\ket{\pm}_i$, with $\ket{\pm} = \frac{1}{\sqrt{2}} (\ket{1} \pm \ket{0})$. When the measurement rate passes a critical value $p_{\#}$, we observe that the total charge fluctuations $\langle\delta^2{\cal Q}\rangle$ diminish in a finite time. In the charge-fuzzy phase, however, when $p < p_{\#}$, the fluctuations $\langle\delta^2{\cal Q}\rangle$ remain finite for times smaller than a length-dependent timescale $t_{\#}$, which diverges as $L \to \infty$. In fact, charge fluctuations provide a very convenient tool to investigate the transition from the fuzzy phase of charge to the charge-sharp phase. In the absence of measurement, the fluctuations remain at the same maximal value $\langle\delta^2{\cal Q}\rangle = L/4$, corresponding to the initial state, and, in fact, a fully mixed state with equal probabilities of $\ket{0}$ and $\ket{1}$ for any qubit. By turning the monitoring on and increasing the measurement rate, the fluctuations at timescales $t \gtrsim L$ start to decrease and eventually become fully suppressed as we approach the charge-sharp phase.

\begin{figure}[t!]
\centering
\includegraphics[width=0.9\linewidth]{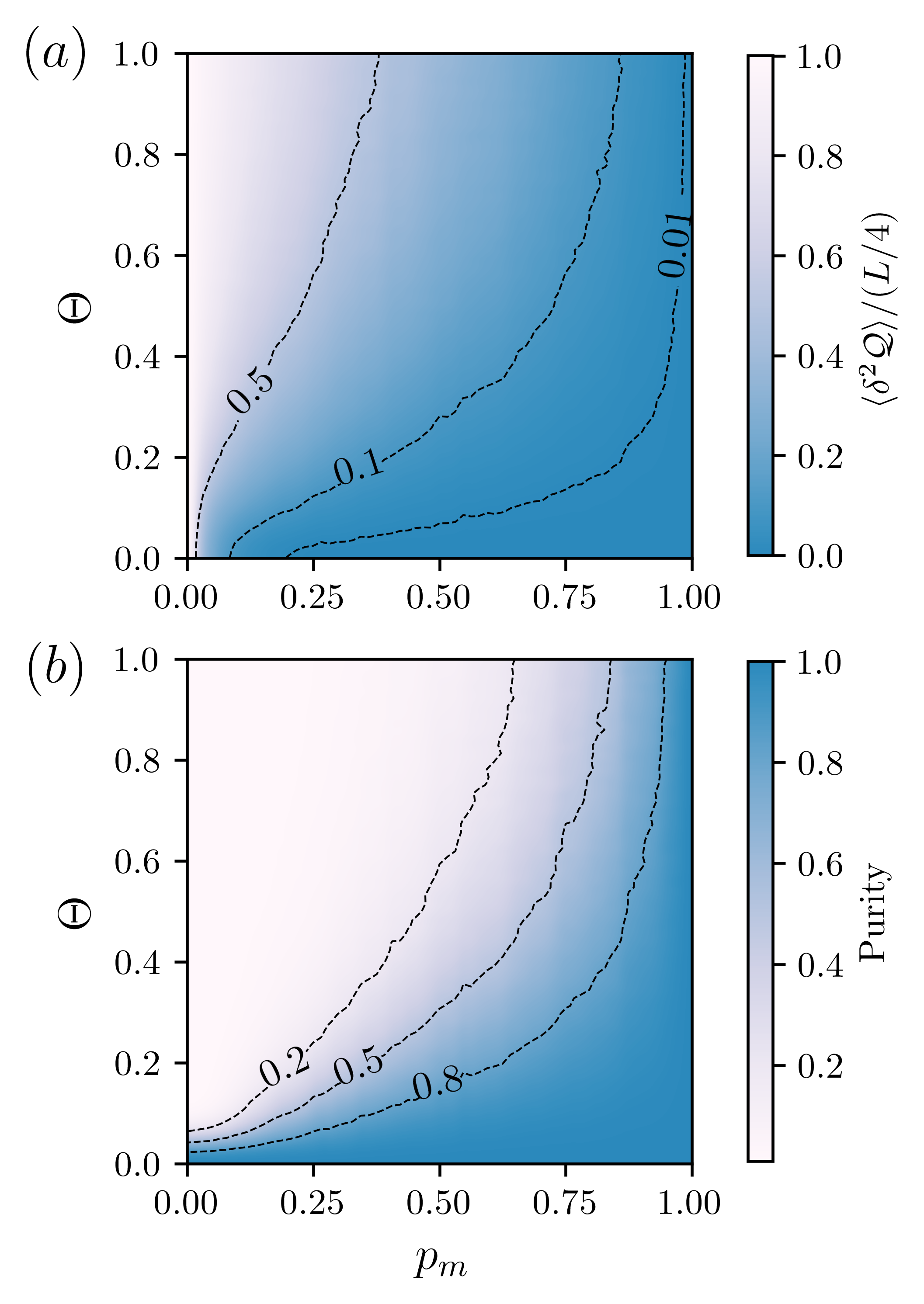}
\caption{\label{fig4} 
\textbf{Effect of noise in $U(1)$ monitored circuits.}
\textbf{(a)} Total charge fluctuations in a $U(1)$-symmetric circuit as a function of measurement rate $p_{\rm m}$ and symmetry-breaking noise amplitude $\Theta$. \textbf{(b)} Same diagram for purity. The results are for circuits of length $L=12$, averaged over $N_{\cal R} = 100$ noise realizations and $N_{\cal U,M} = 200$ realizations for unitary sets and measurement outcomes.}
\end{figure}

\emph{\color{blue} Charge-sharpening with noise}.---Similar to the effect in adaptive circuits, noise explicitly breaks the symmetry of $U(1)$ circuits, thus hindering the transition to a charge-sharp state. However, by increasing the measurement rate, the charge-sharp states begin to re-establish despite the presence of noise. This result is illustrated in Fig.~\ref{fig4}(a), which shows the scaled steady-state value of charge fluctuations as a function of measurement rate and noise amplitude for a circuit with length $L=12$ and depth $T=2L$. Although the transition between fuzzy and charge-sharp states cannot be precisely identified from a single length calculation, it is evident that, at any noise amplitude, increasing the measurement rate eventually leads to very low fluctuations, indicating the onset of the charge-sharp phase. This finding is intriguing because, unlike adaptive circuits, there is no corrective feedback mechanism to directly counteract the mixing effect of noise.

The primary reason for the noise resilience in $U(1)$ circuits---where increasing the measurement rate substantially mitigates the destructive effects of noise---is similar to the underlying mechanism for the very existence of charge-sharpening effect itself. In the standard noise-free scenario, when the circuit starts from a highly mixed (fuzzy charge) state, subsequent measurements progressively narrow the charge distribution. This charge-sharpening process becomes particularly effective beyond a certain measurement rate $p_{\#}$. In the presence of noise, additional fuzziness is introduced into the charge distribution. However, similar to the noise-free case, measurements counteract the broadened charge distribution, mitigating the initial state’s fuzziness and errors at sufficiently high measurement rates. As noted earlier, while the exact transition points cannot be determined from a single size simulation, the numerical results in Fig.~\ref{fig4}(a) qualitatively demonstrate that the critical measurement rate $p_{\#}(\Theta)$ for charge sharpening increases with the noise amplitude $\Theta$.

We find that increasing the measurement rate not only re-establishes a charge-sharp state in a noisy circuit, but can also eventually lead to purification. This is demonstrated in Fig.~\ref{fig4}(b), where the purity ${\rm Tr}(\rho^2)$ is plotted as a function of both measurement rate and noise amplitude. In our implementation, noise is the only source of mixedness, as the circuit begins with a pure initial state. Therefore, in the absence of noise ($\Theta = 0$), the system remains pure at all measurement rates. However, a finite noise results in a mixed state especially at lower measurement rates. By increasing the measurement rate and post-selecting on the measurement outcomes, the purity improves, and nearly pure states are obtained at very high $p_{\rm m}$, even in the presence of strong noise. Similar to adaptive circuits, in the presence of noise, the abrupt transition between phases is smoothed, leading to a gradual crossover between the fuzzy and charge-sharp phases as the measurement rate is varied.

A crucial point here is the necessity of postselecting measurement outcomes to first purify the state within a given charge sector and then to refine it to a specific state within that sector. These steps are essential for achieving the charge-sharpening and purification transitions. This important aspect differentiates our noisy $U(1)$ model from the noisy adaptive circuits discussed in a previous section, which are inherently postselection-free. However, it is important to note that we only need to postselect over measurement outcomes, without artificially postselecting noise realizations. In fact, for noisy $U(1)$-symmetric model, the density matrix is calculated 
for each set of noise-free parts of the unitaries and measurement outcomes as
\begin{align}
    \rho_{\cal U,M}(t) &= \mathbbm{E}_{\cal R} 
(\ket{\Psi_{\cal R,U,M}(t)}\bra{\Psi_{\cal R,U,M}(t)}),
\label{eq:rho_U1}
\end{align}
by only taking the noise average. 
Accordingly, the average quantities are
calculated using
\begin{align}\label{eq:average-taking-U1}
    \langle A \rangle = {\mathbbm E}_{\cal U,M} 
    \left[{\rm Tr}(\rho_{\cal U,M}\,A) \right],
\end{align}
corresponding to postselection over measurement outcomes and non-noisy part of the unitaries. 
This highlights a significant aspect of our model: it accommodates a realistic noisy circuit where there is no control over the noise across different circuit executions.

\begin{figure}[t!]
\centering
\includegraphics[width=0.9\linewidth]{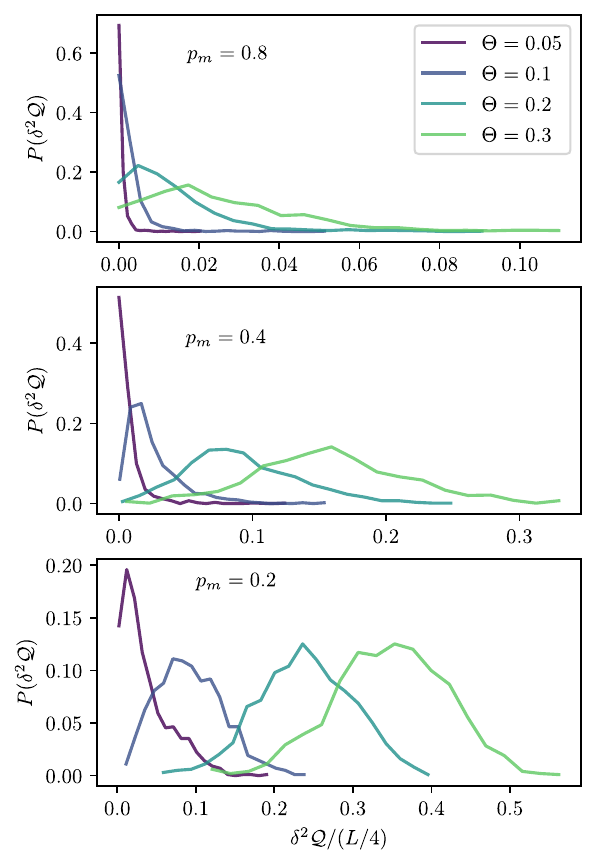}
\caption{\label{fig5} 
\textbf{
Distribution of charge fluctuations in noisy U(1) circuits.} 
The three panels correspond to different measurement rates, $p_{\rm m}$, and show the normalized distribution of charge fluctuations, $P(\delta^2 {\cal Q})$, for some noise amplitudes $\Theta$. For very weak noise, all panels depict a charge-sharp state, characterized by a highly peaked and narrow distribution $P(\delta^2 {\cal Q})$ centered around zero. As the noise increases, the distribution broadens, and its peak shifts toward higher fluctuation values, with the shift becoming more pronounced as $p_{\rm m}$ decreases. The system length is $L=12$, and the
distributions are obtained from histograms of charge fluctuations across $N_{\cal U, M}=1000$ realizations each averaged over $N_{\cal R} = 100$ noise realizations.
}
\end{figure}

We also examine the realization-resolved distribution of charge fluctuations, as shown in Fig.~\ref{fig5}, for different measurement rates $p_{\rm m}$, and a fixed noise amplitude $\Theta$. These distributions are generated from histograms of charge fluctuations across many realizations of the unitary operations and measurement outcomes sets, $\{{\cal U, M}\}$, with each realization averaged over noisy circuit runs ($N_{\cal R} =100$). The histograms are divided into 20 bins, covering the range of fluctuations with nonzero probability for different values of $p_{\rm m}$ and $\Theta$. In the absence of noise, all panels correspond to a charge-sharp state with a narrow, sharply peaked distribution around zero, as $p_{\rm m} > p_\#$. As noise is introduced, the distribution broadens, and its peak shifts toward higher fluctuation values. However, this shift is minimal for a high measurement rate ($p_{\rm m} = 0.8$), where all realizations exhibit very small fluctuations, characteristic of a charge-sharp state. As $p_{\rm m}$ decreases, the broadening and shifting of the charge fluctuation distribution become more pronounced, eventually leading to a fuzzy phase when the distribution centers around a finite value of $0\lesssim\delta^2 {\cal Q} \lesssim L/4$. This indicates a gradual, continuous change in the distribution as noise is varied, signaling a crossover between the charge-sharp and fuzzy phases rather than a sharp phase transition. This behavior contrasts with the noise-free case, where an almost abrupt transition occurs between the two phases as a function of $p_{\rm m}$. Notably, near the transition point ($p_{\rm m} \sim p_\#$), the distribution exhibits a bimodal form, distinguishing it from both the fuzzy and charge-sharp phases~\cite{Agrawal_2022}. We observe that starting from a charge-sharp state ($p_{\rm m} > p_\#$) and introducing noise causes the distribution $P(\delta^2 {\cal Q})$ to remain unimodal, gradually and continuously interpolating between the charge-sharp and fuzzy phases.

\section{Discussion}
The effect of noise and decoherence on entanglement and the associated measurement-induced phase transition has been studied extensively~\cite{Altman2022open,Liu2024PRL,liu2024noise}. There have also been theoretical proposals to protect the entanglement phase transition from decoherence noise using schemes that apply quantum-enhanced operations~\cite{qian2024protect}. In this scheme, instead of performing projective measurements, quantum information is transferred and stored in a quantum memory for post-processing by a quantum computer~\cite{kelly2024}.

In this work, we have addressed the symmetry-breaking role of noise on absorbing-state and sharpening transitions in adaptive and symmetric quantum circuits. Our numerical results show that the noise induced by generic single-qubit errors only deteriorates the absorbing-state and charge-sharp phases at low and intermediate measurement rates. However, we find that at sufficiently high measurement rates, the destructive effect of noise, even at maximal error rates, can be suppressed, thus retrieving the absorbing-state and charge-sharp phases. The underlying reasons for the noise resilience and re-establishment of these phases in the two considered types of circuits are not exactly the same. In adaptive circuits, the feedback mechanism following mid-circuit measurements helps to cancel the noise effects. In contrast, in symmetric circuits where there is no corrective feedback, postselecting over the measurement outcomes is the key tool in combating noise.

Another striking finding is that a generic single-qubit gate error, bounded by a maximum angle $\theta_{\rm max}\equiv \pi \Theta$, results in partially coherent noise. The coherent contribution, determined by the average angle $\phi = \theta_{\rm max}/2$, qualitatively alters the dynamics, revealing a fully quantum character. 
This is in sharp contrast to noiseless case where the quantum channel describing the averaged effects of measurement and unitary gates is found to be effectively classical~\cite{Khemani2022Absorbing}.
As a result of inherent non-classicality in the presence of noise, the system exhibits a crossover rather than a sharp transition between the fuzzy and ordered (charge-sharpened or absorbing) phases.

Quantum circuits with feedback are currently available in various quantum computing architectures~\cite{Monroe_measurement_2022, experiment2023_feedback_ion_trap, devoret_2023_feedback}. There is no major bottleneck for achieving noise resilience in adaptive circuits even with existing NISQ hardware. However, postselection with an exponential execution overhead~\cite{footnote4} is only feasible for small circuits with $L \sim 10$ qubits. Given that there have already been various attempts and proposals to bypass the postselection problem in noise-free circuits~\cite{XEB-altman, XEB-experiment, Gullans2023neural, Ippoliti_postselection, poyhonen2024}, an interesting problem to explore is finding alternative efficient methods that do not require full postselection but still enable the observation of charge-sharpening transitions even in noisy settings.

Based on the availability of feedback mechanisms in existing quantum hardware, not only can our results be directly tested, but this approach may also offer potential side applications in noise characterization.
In the standard method known as \emph{randomized benchmarking} techniques, the error rates and fidelities of quantum gates are typically characterized by applying long sequences of randomly sampled gates, often from the Clifford group~\cite{Knill2008,Gambetta2011,Eisert2022}. Based on our findings, we propose an alternative approach that leverages the relationship between the error amplitude ($\Theta$) or gate fidelity ($\mathbbm F$) and the average value of the absorbing-state order parameter ($\bar n$) in adaptive circuits. This connection allows us to estimate the average gate fidelity of an otherwise symmetric circuit by measuring the steady-value of an order parameter, avoiding the need for complex gate sequences. This method for characterizing the average gate fidelity can provide sufficient information about error rates averaged over all of the qubits in scenarios where we are primarily interested in qubit-averaged quantities, such as in quantum simulations. This approach is a practical alternative to conventional randomized benchmarking when the focus is on system-wide averages rather than individual qubit performance.

\section{Methods}
In our numerical calculations, we first consider a sufficiently large set of individual trajectory dynamics that involve the application of two-qubit unitary gates, random measurement processes, and, in the case of adaptive circuits, additional corrective single-qubit unitaries following the measurements. For each trajectory, the noise-free part of the unitaries is sampled uniformly from a symmetry-constrained random Haar set of two-qubit operators. The measurement sites are randomly chosen with a probability $p_{\rm m} $ at each time step, and the outcomes of each single-qubit measurement are drawn from a two-state distribution based on the Born probabilities of the states $ \ket{0} $ and $ \ket{1} $.
The noise terms $R^j_{\alpha}(\theta_j)$ are implemented according to their description in the main text by randomly selecting a set of qubits and applying a random rotation with angle $\theta_j$ and a rotation axis $ \alpha \in \{x, y, z\} $. For each run and at each time instance, any number of qubits between zero and $L$ can be chosen, resulting in an average of $L/2$ noisy qubits in the limit of large system sizes and long times. After executing each trajectory (i.e., each circuit execution), we construct a density matrix $ \rho $ from the final states according to Eqs. \eqref{eq:rho_adaptive} and \eqref{eq:rho_U1}, for adaptive and $U(1) $-charge conserving cases, respectively.

In the $U(1)$-charge conserving case, post-selection of measurement outcomes is necessary. Without post-selection, the combination of noise and random measurement outcomes would result in a completely mixed state. Therefore, post-selection is essential to diagnose the charge-sharp state and assess purification when noise is present. In this setting, we construct the density matrix for each sequence of measurement outcomes separately. The primary difference between the implementation of adaptive versus $U(1)$-symmetric circuits, aside from the form of the unitaries and the presence/absence of feedback, is that in the $U(1)$-symmetric case, the average over measurement outcomes is taken after obtaining the density matrix for a given sequence of outcomes, as described in Eq. \eqref{eq:average-taking-U1}. This corresponds to post-selecting on measurement outcomes in experiments.

For smaller circuits ($L=10,12$), we numerically evaluate the application of the unitaries in a brickwork structure, as well as the measurements and noise terms, on the multi-qubit state $ \ket{\Psi}$, using our Python code. For larger circuits, where the measurement rate is sufficiently high (as in the results shown in Fig.~\ref{fig3}), we use the package PastaQ~\cite{pastaq} for MPS-based calculations.

The analytical results concerning the following topics are presented in three separate appendices: (a) the relationship between the noise amplitude 
$\Theta$ and the average gate fidelity $\mathbbm{F}$, (b) the decomposition of the single-qubit error channel into its coherent and incoherent contributions, and (c) the interplay between the measurement and noise channels, along with the classical approximation that arises from neglecting the coherent contributions encoded in the commutator of the single-qubit density matrix and Pauli matrices.

\vspace{0.5cm}
\emph{\color{blue} Acknowledgement}.---
M.N.I acknowledges the support by the European Union and the
European Innovation Council through the Horizon Europe
project QRC-4-ESP (Grant Agreement no. 101129663) and
the Academy of Finland through its QTF Center of Excellence program (Project No. 312298).
T.O. and A.G.M. acknowledge Jane and Aatos Erkko
Foundation for financial support. T.O. also acknowledges
the Finnish Research Council project 362573. 

\appendix

\section{Average gate fidelity in presence of noise}\label{app:gate-fidelity}
In this section, we derive the relationship between the noise amplitude $\Theta = \theta_{\rm max}/\pi$, and the average gate fidelity. The average gate fidelity for a generic noisy quantum channel is defined as
\begin{equation}
    {\mathbb F}_{\rm ave} = \int dV_\psi \bra{\psi} \mathcal{E}(\ket{\psi}\bra{\psi}) \ket{\psi},
\end{equation}
which is the average of the gate fidelity 
over the entire Hilbert space of a qubit involved in the gate operation.
As outlined in the main text, the noise axis $\alpha = x, y, z$ is chosen randomly, with each direction having a probability of $\frac{1}{3}$. The noise angle $\theta$ is drawn from a uniform random distribution over the range $[0, \theta_{\rm max}]$. Therefore, for the single-qubit errors considered in this context, the corresponding noisy quantum channel can be expressed as 
\begin{equation} 
\mathcal{E}^{\rm err}(\ket{\psi}\bra{\psi}) = \frac{1}{3} \int_0^{\theta_{\rm max}} \frac{d\theta}{\theta_{\rm max}} \sum_{i} R_\alpha(\theta)(\ket{\psi}\bra{\psi})R_i^\dagger(\theta), \end{equation} 
where $R_\alpha(\theta)$ represents the noise component of the applied quantum gates.
The normalization of the quantum channel can be easily verified by noticing
\begin{align}
 \frac{1}{3} \int_0^{\theta_{\rm max}} \frac{d\theta}{\theta_{\rm max}} \sum_{\alpha=x,y,z} R_i(\theta) R_i^\dagger(\theta) =    \frac{3}{3} \int_0^{\theta_{\rm max}} \frac{d\theta}{\theta_{\rm max}} \mathbbm{1}=\mathbbm{1}
\end{align}
More generically this channel is combined
with the identity (noiseless) channel, $\mathcal{E}^{\rm ide}(\rho)=\rho$, to take into account the finite probability $1-\gamma$ of not having error in applying gates, which yields
\begin{align}\label{eq:noise+Id}
  \mathcal{E}^{\rm noisy}(\rho) =   \gamma \mathcal{E}^{\rm err}(\rho) + (1-\gamma)\rho.
\end{align}
Considering this form of noisy quantum channel, and after some straightforward algebra, we obtain
\begin{align}
    {\mathbb F}_{\rm ave}  & = 1-\gamma+\frac{\gamma}{3} 
    \sum_{i}
    \int_0^{\theta_{\rm max}} \frac{d\theta}{\theta_{\rm max}} 
    \int dV_\psi 
    |\bra{\psi} R_i(\theta) \ket{\psi}|^2  \nonumber \\
    & = 1-\gamma+\frac{\gamma}{3}  \sum_{i} \int \frac{d\theta}{\theta_{\rm max}} \frac{1}{6}
    \left[2 + |{\rm Tr} R_\alpha(\theta)|^2 \right]\nonumber \\
    & =  1-\gamma+\gamma
    \int_0^{\theta_{\rm max}} \frac{d\theta}{\theta_{\rm max}} 
    \frac{1 + 2\cos^2 (\theta/2)}{3}\nonumber \\
    & =1-\frac{\gamma}{3} \left( 1 - \frac{\sin \theta_{\rm max}}{\theta_{\rm max}}\right).
\end{align}
By increasing both the noise amplitude $\theta_{\rm max}$ and its rate $\gamma$, gate fidelity decreases.
In the special case of noise rate $\gamma=1$, the average fidelity of single qubit gates decreases monotonically from 1 to $2/3$
by varying the noise amplitude $\theta_{\rm max}$ between $0$ and $1$.
Throughout our numerical calculations, we set $\gamma=1/2$, which corresponds to half of the qubits, on average, being affected by noise at each time step.

\section{Decomposition of error channel into coherent and incoherent parts}\label{app-channels}

Here, we derive an expression for the quantum channel in terms of Pauli operators. Using the expression for the rotation operator, $R_{\alpha}(\theta) = \cos \frac{\theta}{2}  {\mathbbm{1}} + i \sin \frac{\theta}{2} \sigma_{\alpha}$, it is straightforward to show that
\begin{align} \label{eq:rotation-to-pauli}
R_{\alpha}(\theta) \rho R_{\alpha}^\dagger(\theta) &= \cos^2\left(\frac{\theta}{2}\right) \rho + \sin^2\left(\frac{\theta}{2}\right) \sigma_{\alpha} \rho \sigma_{\alpha} \nonumber \\
&\quad + \frac{i}{2} \sin \theta [\sigma_{\alpha}, \rho].
\end{align}
The rotation angle $\theta$ is sampled from a uniform distribution within the interval $[0, \theta_{\rm max}]$, where $\theta_{\rm max}$ represents the maximum noise angle. Note that in the main text, the substitution $\theta_{\rm max} \to \pi \Theta$ has been made such that $\Theta \in [0,1]$. After integrating over $\theta$, the error channel is given by
\begin{align} \label{eq-error-channel}
\mathcal{E}^{\rm err}(\rho) &= \frac{1}{3} \sum_{{\alpha}=x,y,z} \left[ \left( \frac{1}{2} + \frac{\sin \theta_{\rm max}}{2 \theta_{\rm max}} \right) \rho \right. \nonumber \\
&\quad \left. + \left( \frac{1}{2} - \frac{\sin \theta_{\rm max}}{2 \theta_{\rm max}} \right) \sigma_{\alpha} \rho \sigma_{\alpha} \right. \nonumber \\
&\quad \left. + i \frac{1 - \cos\theta_{\rm max}}{2 \theta_{\rm max}} [\sigma_{\alpha}, \rho] \right].
\end{align}
It can be readily verified that this map satisfies the requirement of complete-positiveness. In particular, it is important to note that the last term remains Hermitian because the commutator is anti-Hermitian ($\left[\sigma_{\alpha},\rho\right]^\dagger = -\left[\sigma_{\alpha},\rho\right]$), but the imaginary factor $i$ ensures Hermiticity. Additionally, since $\text{Tr}([\sigma_{\alpha}, \rho]) = 0$ and $\text{Tr}(\sigma_{\alpha} \rho \sigma_{\alpha}) = \text{Tr}(\rho)$, the map is trace-preserving. Furthermore, this map is also a \emph{unital channel}, meaning that $\mathcal{E}(\mathbbm{1}) = \mathbbm{1}$.

The presence of the last term in Eq. \eqref{eq-error-channel} suggests that our error channel also has a coherent contribution. In fact, as we will demonstrate, the error channel above can be decomposed into fully coherent and incoherent (Pauli) components. To see this, recall that a fully coherent quantum channel, arising from purely unitary dynamics $\rho \to U\rho U^\dagger$ of a qubit, can always be expressed in terms of rotation channels, given by Eq.~\eqref{eq:rotation-to-pauli}. We denote these three coherent channels as 
\begin{equation}
  {\cal E}^{\rm coh}_{\alpha}(\rho) = R_{\alpha}(\varphi) \rho R_{\alpha}^\dagger(\varphi).  
\end{equation}

On the other hand, a Pauli depolarizing channel can be decomposed into three fundamental dephasing channels:
\begin{align}
    {\cal E}^{\rm incoh}_{\alpha} (\rho) = (1-\eta) \rho + \eta\: \sigma_\alpha \rho \sigma_\alpha.
\end{align}
Let's now construct the combination of coherent and incoherent channels
which leads to
\begin{align} \label{eq:combined}
    {\cal E}^{\rm err} & = \frac{1}{3}\sum_{\alpha} 
     {\cal E}^{\rm incoh}_{\alpha} \circ {\cal E}^{\rm coh}_{\alpha}(\rho) \nonumber\\
     & =\frac{1}{3}\sum_{\alpha} \left\{  (1-{\cal X})\rho + {\cal X}\: \sigma_\alpha\rho\sigma_\alpha + i{\cal Y}\:[\sigma_\alpha,\rho] \right\},  
\end{align}
where
\begin{align}
    {\cal X} & = \eta \cos^2(\frac{\varphi}{2})  + (1-\eta) \sin^2(\frac{\varphi}{2}), \\
    {\cal Y} & = (\frac{1}{2}-\eta) \sin \varphi.
\end{align}
One can readily see that the error channel in Eq. \eqref{eq-error-channel} actually has a similar form to the combined channel given by \eqref{eq:combined} with
\begin{align}
        {\cal X}^{\rm err} & =    \frac{1}{2} - \frac{\sin \theta_{\rm max}}{2 \theta_{\rm max}} 
, \\
    {\cal Y}^{\rm err} & = \frac{1 - \cos\theta_{\rm max}}{2 \theta_{\rm max}}.
\end{align}
By equating the expressions for the channel parameter ${\cal X}$ and $\cal Y$
and after some straightforward algebra we find that the effective coherent rotation angle $\varphi$ and depolarization rate $\eta$ 
in terms of maximum noise angle $\theta_{\rm max}$ given by
\begin{align}
    \varphi = \frac{\theta_{\rm max}}{2}, \qquad \eta = \frac{1}{2} - \frac{\sin(\frac{\theta_{\rm max}}{2})}{\theta_{\rm max}} 
\end{align}

\section{Interplay of measurement and noise from a quantum channel perspective}\label{app-measurement-noise-channel}

The effect of measurement, combined with corrective feedback, on the averaged single-qubit density matrix (without postselecting measurement outcomes) is described by the following quantum channel:
\begin{align}
    {\cal E}^{\rm m+f}(\rho) = (1 - p_{\rm m}) \rho + p_{\rm m} \left( \Pi_{\uparrow} \rho \Pi_{\uparrow} + X \Pi_{\downarrow} \rho \Pi_{\downarrow} X \right),
\end{align}
where $X$ is the correction operator (which flips the two computational basis states), and $\Pi_{\uparrow} = \ket{\uparrow}\bra{\uparrow}$
and $\Pi_{\downarrow} = \ket{\downarrow}\bra{\downarrow}$ are projectors onto the spin-up ($1$) or spin-down ($0$) states, respectively. In the case of non-adaptive circuits (such as the $U(1)$-charge conserving model considered here), the measurement channel is simply:
\begin{align}
    {\cal E}^{\rm m}(\rho) = (1 - p_{\rm m}) \rho + p_{\rm m} \left( \Pi_{\uparrow} \rho \Pi_{\uparrow} + \Pi_{\downarrow} \rho \Pi_{\downarrow} \right),
\end{align}
which describes the averaged effect of measurement without corrective feedback.

These channels can be represented as matrices
\begin{align}
    \Phi({\cal E}^{\rm m+f}) &= \begin{pmatrix}
    1 & 0 & 0 & p_{\rm m} \\
    0 & 1 - p_{\rm m} & 0 & 0 \\
    0 & 0 & 1 - p_{\rm m} & 0 \\
    0 & 0 & 0 & 1 - p_{\rm m} \\
    \end{pmatrix},\\    
    \Phi({\cal E}^{\rm m}) &= \begin{pmatrix}
    1 & 0 & 0 & 0 \\
    0 & 1 - p_{\rm m} & 0 & 0 \\
    0 & 0 & 1 - p_{\rm m} & 0 \\
    0 & 0 & 0 & 1 \\
    \end{pmatrix},
\end{align}
in the doubled Hilbert space 
acting over the density matrix of the single qubit recast as a column vector according to the basis $\{\ket{\uparrow}\bra{\uparrow}, \ket{\uparrow}\bra{\downarrow}, \ket{\downarrow}\bra{\uparrow}, \ket{\downarrow}\bra{\downarrow}\}$.
The matrix form clearly shows that the computational basis measurement does not couple the off-diagonal and diagonal elements of the single-qubit averaged density matrix. Similarly, as discussed in Ref.~\cite{Khemani2022Absorbing}, the averaged effect of noiseless two-qubit unitaries only involves the diagonal elements in the two-qubit basis, as all the terms coupling off-diagonal and diagonal terms have phase factors which sum to zero by taking the average over infinitely many realization of random unitaries. Consequently, the dynamics of the diagonal elements of the averaged density matrix can be mapped to a classical stochastic process.

By introducing noise, particularly due to the coherent contributions described by the channels ${\cal E}^{\rm coh}_{\alpha}(\rho)$, the classical mapping no longer holds. This breakdown occurs because, upon decomposing the channels ${\cal E}^{\rm coh}_{\alpha}(\rho)$ according to Eq. \eqref{eq:rotation-to-pauli}, the commutator terms involving the Pauli matrices $X$ and $Y$ with $\rho$ mix the off-diagonal and diagonal components of the density matrix. This mixing becomes evident when examining the corresponding $4\times4$ matrix representations of these commutator terms within the channels:
\begin{align}
    \Phi{(\rho \to [X,\rho])} &= \begin{pmatrix}
    0 & -1 & 1 & 0 \\
    -1 & 0 & 0 & 1 \\
    1 & 0 & 0 & -1 \\
    0 & 1 & -1 & 0 \\
    \end{pmatrix},\\    
    \Phi{(\rho \to [Y,\rho])} &= i\begin{pmatrix}
    0 & -1 & -1 & 0 \\
    1 & 0 & 0 & -1 \\
    1 & 0 & 0 & -1 \\
    0 & 1 & 1 & 0 \\
    \end{pmatrix}.
\end{align}
This highlights an important characteristic of the coherent part of the noise channel, which introduces dynamics that are inherently quantum mechanical, without any possibility of being mapped onto a classical stochastic model, in contrast to the noise-free case studied in~\cite{Khemani2022Absorbing}.

Having established the non-classical nature of the noise due to partially coherent contributions, we now proceed with a further simplification by neglecting the commutator terms $[\sigma_\alpha, \rho]$ in the error channel \eqref{eq-error-channel}, which incorporate the coherent contribution. This simplification can be interpreted as a classical approximation $\tilde{\mathcal{E}}^{\rm err}$ to the noise channel:
\begin{align}
    \mathcal{E}^{\rm err}(\rho) 
    \approx 
    \tilde{\mathcal{E}}^{\rm err}(\rho) 
    = (1-\nu)\rho + \frac{\nu}{3} \sum_{\alpha=x,y,z} \sigma_{\alpha} \rho \sigma_{\alpha},
\end{align}
where 
\begin{align}
    \nu = \frac{1}{2} \left( 1 - \frac{\sin \theta_{\rm max}}{\theta_{\rm max}} \right).
\end{align}
More generally, considering partially noisy situations with a finite probability $\gamma$ of occurring error as in Eq. \eqref{eq:noise+Id}, the same form for classical approximate channel holds with the modification
\begin{align}
    \nu \to {\nu}' =\gamma\nu = \frac{\gamma}{2} \left( 1 - \frac{\sin \theta_{\rm max}}{\theta_{\rm max}} \right).
\end{align}

The combination of this approximate noise channel with the measurement process leads to the following form for the combined channels:
\begin{align}
    {\cal E}^{\rm m+f} \circ \tilde{\mathcal{E}}^{\rm noisy} (\rho) &= (1 - p_{\rm m}) 
    \Big[ 
    (1-\nu')\rho +
    \frac{\nu'}{3} \sum_{\alpha} \sigma_\alpha \rho \sigma_\alpha
    \Big] \nonumber \\
    &+ p_{\rm m} \left( \Pi_{\uparrow} \rho \Pi_{\uparrow} + X \Pi_{\downarrow} \rho \Pi_{\downarrow} X \right),
\end{align}
which describes noisy adaptive circuits.

One can easily verify that this expression effectively corresponds to a classical model, without any coupling between the diagonal and off-diagonal elements of the single-qubit density matrix $\rho$. By focusing on the part that only involves the diagonal elements, the channel can be represented as:
\begin{align}
    \begin{pmatrix}
        \rho_{\uparrow\uparrow} \\
        \rho_{\downarrow\downarrow}
    \end{pmatrix} \to
    \begin{pmatrix}
        1-\xi & p_{\rm m} + \xi \\
        \xi & (1-p_{\rm m}) - \xi
    \end{pmatrix}
    \begin{pmatrix}
        \rho_{\uparrow\uparrow} \\
        \rho_{\downarrow\downarrow}
    \end{pmatrix}, \label{eq:combined-classical-err-meas}
\end{align}
where, for the sake of compactness, we have defined a new parameter
\begin{align}
    \xi = \frac{2}{3} \nu' (1 - p_{\rm m}) = \frac{\gamma}{3} (1 - p_{\rm m}) \left( 1 - \frac{\sin \theta_{\rm max}}{\theta_{\rm max}} \right).
\end{align}

\begin{figure}[t!]
\centering
\includegraphics[width=0.9\linewidth]{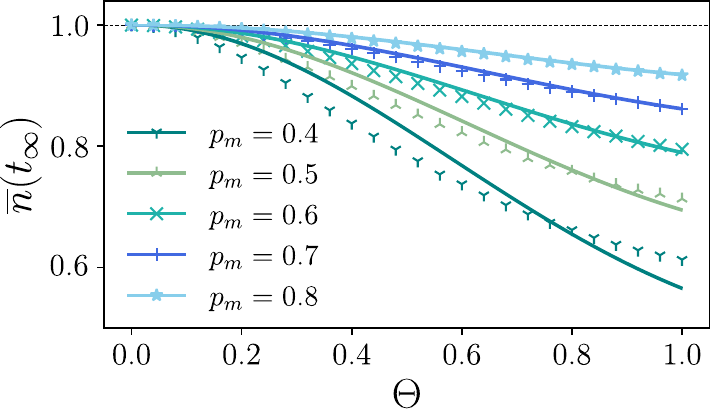}
\caption{\label{fig6} 
\textbf{Noisy absorbing-state order parameter.} 
Comparison between the numerical results shown in Fig.~\ref{fig3} and the classical channel model for the steady-state value of the order parameter, given by Eq.~\eqref{eq:classical-order-parameter}. We set $\gamma = 1/2$, consistent with the numerical simulations. The approximate model shows better agreement with the numerical results at high measurement rates. However, for intermediate measurement rates, deviations between the approximate model and the numerical simulations become evident (see text for more details).
}
\end{figure}

The steady-state behavior corresponding to this approximate classical dynamics involving measurement, feedback, and incoherent noise can be determined from the equation:
\begin{equation}
    \begin{pmatrix}
        \rho_{\uparrow\uparrow} \\
        \rho_{\downarrow\downarrow}
    \end{pmatrix}_{\rm st.} =
    \begin{pmatrix}
        1-\xi & p_{\rm m} + \xi \\
        \xi & (1-p_{\rm m}) - \xi
    \end{pmatrix}
    \begin{pmatrix}
        \rho_{\uparrow\uparrow} \\
        \rho_{\downarrow\downarrow}
    \end{pmatrix}_{\rm st.},
\end{equation}
with the solution obtained by solving this system. Using this result, the steady-state value of the order parameter from this classical picture can be approximated as:
\begin{align} 
    \label{eq:classical-order-parameter}
    \bar{n} = \rho_{\uparrow\uparrow} - \rho_{\downarrow\downarrow} 
    \approx \frac{p_{\rm m}}{\sqrt{p_{\rm m}^2 + 2 \xi \left(p_{\rm m} + \xi\right)}}.
\end{align}

The dependence of the order parameter on the noise amplitude, defined as $\Theta \equiv \theta_{\rm max}/\pi$, for different measurement rates, as obtained from the approximate classical model in Eq. \eqref{eq:classical-order-parameter}, is shown in Fig.~\ref{fig6}. To assess the accuracy of this approximate model, we also re-plot the numerical results obtained using the matrix product state technique, previously shown in Fig.~\ref{fig3}(b). For high measurement rates ($p_{\rm m} \gtrsim 0.6$), the classical model exhibits reasonable agreement with the numerical results from the full quantum calculations. However, as the measurement rate decreases, the disparity between the classical approximation and numerical results becomes more pronounced. This suggests that, at relatively low measurement rates and for intermediate to strong noise levels, the coherent part of the noise plays a more significant role in the dynamics.

\bibliography{refs}

\begin{thebibliography}{64}
\bibitem [{foo({\natexlab{a}})}]{footnote1}%
  \BibitemOpen
  \bibinfo {note} {Note that this also includes especial cases when, for instance, only one of the qubits in the pair becomes noisy with \(V^{j,j+1} \) being either \(R_k^{j}(\theta_j)\otimes I^{j+1}\) or \(I^{j}\otimes R^{j+1}(\theta_{j+1})\).}\BibitemShut {Stop}%
\bibitem [{foo({\natexlab{c}})}]{footnote3}%
  \BibitemOpen
  \bibinfo {note} {{As elucidated in various works, the purification transition is closely related to, and typically coincides with, the transition between volume-law and area-law entanglements.}}\BibitemShut {Stop}%
\bibitem [{foo({\natexlab{d}})}]{footnote4}%
  \BibitemOpen
  \bibinfo {note} {{Exponentially large numbers of circuit executions are required to postselect a given set of measurement outcomes.}}\BibitemShut {Stop}%
\end{thebibliography}

\end{document}